\documentstyle[12pt,epsfig]{article}
\def\beq{\begin{equation}}
\def\eeq{\end{equation}}
\def\bea{\begin{eqnarray}}
\def\eea{\end{eqnarray}}
\def\bq{\begin{quote}}
\def\eq{\end{quote}}

\def\bq{\begin{quote}}
\def\eq{\end{quote}}

\def\mpl{\ifmmode \overline M_{P}\else $\overline M_{P}$\fi}
\begin{document} 
\vspace*{-1in} 
\renewcommand{\thefootnote}{\fnsymbol{footnote}} 
\begin{flushright} 
CERN-TH/2000-275 \\
IFT-P082/2000\\
IITK-PHY/2000/20 \\
TIFR/TH/00-54\\
{\bf hep-ph/0010010} \\ 
\end{flushright} 
\vskip 65pt 
\begin{center} 
{\LARGE \bf 
         Extra Dimensions: A View from the Top}\\
\vspace{8mm} 
{\large\bf 
         Smaragda Lola${}^{1}$\footnote{Magda.Lola@cern.ch}
         Prakash Mathews${}^{2}$\footnote{mathews@it.unesp.br}, \\
         Sreerup Raychaudhuri${}^3$\footnote{sreerup@iitk.ac.in, 
                                             sreerup@mail.cern.ch},   
         K.~Sridhar${}^4$\footnote{sridhar@theory.tifr.res.in}
}\\ 
\vspace{10pt} 
{\footnotesize\sf 1) Theory Division, CERN, 
                     CH-1211 Geneva 23, Switzerland.\\

                  2) Instituto de Fisica Teorica, 
                     Universidade Estadual Paulista, \\
                     Rua Pamplona 145, 01405-900, 
                     Sao Paulo, SP, Brazil.\\

                  3) Department of Physics, 
                     Indian Institute of Technology,\\ 
                     Kanpur 208 016, India.

                  4) Department of Theoretical Physics, \\
                     Tata Institute of Fundamental Research,\\  
                     Homi Bhabha Road, 
                     Bombay 400 005, India. } 

\normalsize
 
\vspace{20pt} 
{\bf ABSTRACT:} 
\end{center} 

\noindent
In models with compact extra dimensions, where the Standard Model fields 
are confined to a 3+1 dimensional hyperplane, the $t \bar t$ production
cross-section at a hadron collider can receive significant contributions
from multiple exchange of KK modes of the graviton. These are carefully 
computed 
in the well-known ADD and RS scenarios, taking the energy dependence of 
the sum 
over graviton propagators into account. Using data from Run-I of the 
Tevatron, 95\% C.L. 
bounds on the parameter space of both models are derived. 
For Run-II of the Tevatron and LHC, discovery limits are estimated.

\vskip12pt 
\noindent 
\setcounter{footnote}{0} 
\renewcommand{\thefootnote}{\arabic{footnote}} 
 
\vfill 
\clearpage 

%-------------------------------------------------------------------%
\setcounter{page}{1} 
\pagestyle{plain}

%\section{Introduction}

\noindent In recent years, there has been tremendous interest in higher
dimensional theories where the extra dimensions are compactified in such a
way as to have physically interesting consequences\cite{string}. These
have been inspired by recent developments in our understanding of some
non-perturbative aspects of string theories. The weak-coupling spectrum of
such a theory reveals only one-dimensional strings but, in the strong
coupling regime, $p+1$ dimensional defects embedded in the
higher-dimensional spacetime --- called $D_p$-branes --- arise. These are
dynamical objects analogous to solitons in field theory and are
particularly interesting because it is possible to localize gauge fields
on them. We then have the exciting possibility that the observable
universe is a $D_3$-brane (or 3-brane) with the Standard Model (SM) fields
localized on it, leaving gravity free to propagate in the extra dimensions
or bulk. In such a scenario, it becomes possible to lower the
gravity/string scale to the order of a TeV, without running into
contradiction with experiment. This new TeV-scale physics then provides
fresh perspectives on the hierarchy problem arising in the standard
electroweak theory.

The first realization of these ideas is the ADD scenario proposed by
Arkani-Hamed, Dimopoulos and Dvali\cite{dimo}, where the SM fields are
localized on a 3-brane embedded in a 10-dimensional spacetime consisting
of 4 non-compact Minkowski dimensions, $d$ dimensions compactified with a
large radius $R_c$ and the remaining $6-d$ dimensions compactified to the
Planck length. The Planck scale $M_P$ is then related\cite{dimo} to the
scale $M_S$ of gravitational interactions by $M_P \sim R_c^d M_S^{d+2}$.
Choosing the scale $M_S$ to be of the order of a TeV (which eliminates the
usual hierarchy problem) it follows that $R_c \sim 10^{32/d -19}$~m. This
immediately precludes $d = 1$, but for any value $d > 1$, we find that
$M_S$ can be arranged to be a TeV for compactification radii as large as a
millimetre, without conflicting with gravity experiments\cite{gravexp}.
Effects of non-Newtonian (quantum) gravity can then become apparent at
these surprisingly low values of energy. While it is possible\cite{dimo4}
to construct a physically viable scenario with large $R_c$ which can
survive the existing astrophysical and cosmological constraints, such a
scenario inevitably predicts large deviations from the SM in several
low-energy effects \cite{phenoadd}. Thus, laboratory data can be used to
derive \cite{phenoadd} bounds on $M_S$ in the TeV range --- though some
astrophysical bounds can, in fact, be considerably stronger \cite{cullen}.

The main criticism of the ADD scenario is the reappearance of a large
hierarchy between the string scale $M_S \sim 1$ TeV and the
compactification radius $R_c \sim \left( 10^{-16}~{\rm TeV} \right)^{-1}$,
which makes it difficult to stabilize the size of the large extra
dimensions. This has inspired the RS scheme --- proposed by Randall and
Sundrum \cite{rs} --- where only a single extra dimension is invoked. This
fifth dimension $\phi$ is strongly curved and is compactified on a ${\bf
S}^1/{\bf Z}^2$ orbifold with a radius $R_c$, which is assumed to be
somewhat larger than the Planck length. Two 3-branes are located at the
orbifold fixed points: a `Planck brane' at $\phi=0$ and a `TeV brane' at
$\phi=\pi$. The tension on the branes and the cosmological constant in the
bulk are fine-tuned to give a flat (Minkowski) geometry on the 3-branes.
The bulk metric that guarantees this fine-tuning is 
\begin{equation}
ds^2 = e^{-{\cal K}R_c\phi}\eta_{\mu\nu}dx^{\mu}dx^{\nu}~+~R_c^2d\phi^2 .
\end{equation} 
The novel feature of this metric that it is
{\it non-factorizable} or $warped$ and this leads to interesting consequences,
such as localised graviton fields. The real interest of this model,
however, lies in the fact that the exponential warp factor in the metric
helps to solve the hierarchy problem. Mass scales are determined by the
location $\phi$ of the brane in the extra dimension, since the warp factor
$e^{-{\cal K}R_c\phi}$ acts as a conformal factor for the brane fields,
{\it i.e.}, all masses get rescaled by this factor. This exponential could
be the source of the hierarchy between the electroweak scale and the
Planck scale, because it is possible in this model to generate the 
enormous factor
of $\frac{M_P}{M_{EW}} \sim 10^{15}$ by an argument $\pi{\cal K}R_c$ of
order 30. Physically, the Planck scale is treated as the fundamental scale
in the theory, but the overlap of the graviton wave-function with the TeV
brane on which the gauge particles are localised is very small and this is
what generates the small value of the electroweak scale. There still
remains the less severe problem of stabilizing the compactification radius
$R_c$ against quantum fluctuations, but it has been shown that it is
possible to achieve this, either by introducing an extra scalar field in
the bulk \cite{csaki, gold}, or by invoking supersymmetry \cite{bagger}. 
The bulk
scalar is expected to be light and may have interesting phenomenological
consequences \cite{radion}. A deeper problem is to realize the RS scenario
within the framework of a string theory, and some progress has been made
in this direction \cite{verlinde}.

To determine the consequences of the ADD and RS models for experiments, we
need to extract the low-energy effective theory on the 3-brane where the
SM fields live. We concentrate on the experimental signals of the
Kaluza-Klein excitations of the graviton, which is massless in the bulk,
but, on compactification of the extra dimensions, yields a tower of
massive Kaluza-Klein (KK) excitations on the 3-brane. Using the linearised
gravity approach, the curved metric can be approximated by small
fluctuations $h_{\mu\nu}$ about its Minkowski value. The interactions of
its KK modes $h^{(\vec{n})}_{\mu\nu}$ with the observable particles on the
3-brane are then given by:  
\begin{eqnarray} 
{\rm ADD:}~{\cal L}_{int} & =
& -{1 \over \mpl} \sum_{\vec{n}} T^{\mu\nu}(x) h^{(\vec{n})}_{\mu\nu}(x)
\\ 
{\rm RS:}~~{\cal L}_{int} & = & -{1 \over \mpl} T^{\mu\nu}(x)
h^{(0)}_{\mu\nu}(x) -{e^{\pi {\cal K} R_c} \over \mpl} \sum_n^{\infty}
T^{\mu\nu}(x) h^{(n)}_{\mu\nu}(x) \ , 
\end{eqnarray} 
where
$\mpl=M_P/\sqrt{8\pi}$ is the reduced Planck mass and $T^{\mu\nu}$ is the
symmetric energy-momentum tensor for the observable particles on the
3-brane, computed using the flat space metric. The masses of the
$h^{(\vec{n})}_{\mu\nu}$ are given by 
\begin{eqnarray} 
{\rm ADD:}~M^2_{\vec{n}} & = & \frac{1}{R_c} \left( n_1^2 + n_2^2 + \dots
+ n_d^2 \right)^2 \\ 
{\rm RS:}~~M^2_{n} & = & x_n {\cal K} ~e^{-\pi {\cal K} R_c}
\end{eqnarray} 
where the $x_n$ are the zeros of the Bessel function $J_1(x)$ of order
unity \cite{gold}. It is important to note that the masses are
evenly-spaced in the ADD case, but not so in the RS case.  It is also
obvious that in the ADD scenario, all the $h^{(\vec{n})}$ couple very
weakly to matter (being suppressed by $\mpl^{-1}$), as does the zero-mode
of the KK tower in the RS case. On the other hand the couplings of the
massive RS gravitons are enhanced by the exponential $e^{\pi {\cal K}
R_c}$ leading to interactions of electroweak strength. In the ADD model,
the density of light KK gravitons is very high because of the large value
of $R_c$, and their collective interaction again builds up to electroweak
strength. The Feynman rules in either case are essentially the same as
those worked out\cite{grw, hlz} for the ADD case, except for the overall
warp factor in the RS case.

For the ADD model, then, the only unknown parameter in the theory is the
string scale $M_S$, which is related to the compactification radius $R_c$
by $M_P \sim R_c^d M_S^{d+2}$. For exchange of virtual gravitons, $M_S$
also acts as a cutoff for the sum over KK states. In the RS case, it is
expedient to define the free parameters 
\begin{eqnarray} 
m_0 & = & {\cal K} e^{-\pi {\cal K} R_c} \nonumber \\ 
c_0 & = & {\cal K}/M_P \label{m0c0}
\end{eqnarray} 
where $m_0$ sets the scale for the masses of the excitations, and $c_0$ is
an effective coupling, since the interaction of massive KK gravitons with
matter can be written as 
\begin{equation} 
{\cal L}_{int}  = 
-\sqrt{8\pi} {c_0 \over m_0} \sum_n^{\infty} T^{\mu\nu}(x)
h^{(n)}_{\mu\nu}(x) \ . 
\end{equation} 
Typical values of the parameter $c_0$ lie in the range [0.01, 0.1]. This
estimate results from the requirement that the scale ${\cal K}$ (related
to the curvature of the fifth dimension) is small compared to $\mpl$, but
not too small, since that would introduce a new hierarchy.  Values of
$m_0$ are determined in terms of ${\cal K} R_c \sim 10$, so that $m_0$
ranging from about 50 or 60 GeV to a TeV are consistently obtained.

Before discussing detailed phenomenological consequences of these models,
we take note of some of the generic features. The KK graviton modes in the
ADD model can be very light, with masses ranging from $10^{-13}$ GeV to
0.1 GeV, for $M_S \sim 1$ TeV, depending on the number $d$ of extra
dimensions. For the current generation of high-energy experiments,
performed at energies of order 100 GeV, huge numbers of KK gravitons can
contribute to the same process, leading to effective interactions of
electroweak strength. These interactions have been extensively studied in
the literature \cite{phenoadd}. In the RS model, the individual KK modes are
heavier --- ${\cal O}$(TeV) --- and couple individually with electroweak
strength. One has then, the interesting possibility that they may be
produced on resonance. As one increases the centre-of-mass energy, one may
hope to probe a multi-resonance effect. Moreover, instead of escaping the
detector, as will be the case for ADD gravitons, the massive gravitons
will decay, mostly within the detectors, into pairs of SM particles. Thus,
the experimental manifestations of the RS model are expected to be
somewhat different from those in the ADD model. Some work on the collider
phenomenology of this model has already appeared in the literature. These
include the resonant production of the KK excitations and the virtual
effects in processes like dilepton production at hadron colliders
\cite{dhr}, in deep-inelastic scattering at HERA \cite{drs} and in $e^-
e^-$\cite{gr} and $e\gamma$ colliders. Novel effects like probing strong
gravity via black-hole production at low energies have also been discussed
in the context of the RS model \cite{giddings}.

In this letter, we study, within the framework of the ADD and RS models,
the virtual effects of the exchange of spin-2 KK modes in the production
of $t \bar{t}$ pairs at hadron colliders, in particular, the Tevatron and
LHC. Hadroproduction of $t \bar t$ pairs has already been used to
constrain the ADD model earlier\cite{ours1}, using a low-energy
approximation. We improve upon the earlier results for the ADD model by
using a better approximation, and go on to derive constraints from the $t
\bar t$ cross-section on the parameter space of the RS model, which has
not been done before. 

To estimate the cross-section for $t \bar t$-hadroproduction, in addition
to the SM processes, we consider production mechanisms with $s$-channel
exchange of KK gravitons in $q \bar q$- or $gg$-initiated sub-processes.
The graviton contributions to the sub-process cross-section in the two
models can be written in the common form, as
\begin{eqnarray} 
{d\hat \sigma \over d\hat t}(q \bar q \rightarrow t \bar t) 
& = & {d\sigma \over d\hat t}_{\rm SM} (q \bar q \rightarrow t \bar t) \\
&& + {\pi |C(x_s)|^2 \over 64 \hat s^2 }
\biggl \lbrack 5 \hat s^2(\hat t- \hat u)^2+4 (\hat t-\hat u)^4 
+8 \hat s (\hat t- \hat u)^2 (\hat t+\hat u)-2 \hat s^3 (\hat t+\hat u) -\hat 
s^4\biggr\rbrack ~, \nonumber 
\end{eqnarray} 
assuming massless quarks in the initial state, and
\begin{eqnarray} 
{d\hat \sigma \over d\hat t}(gg \rightarrow t \bar t) 
& = & {d\sigma \over d\hat t}_{\rm SM} (gg \rightarrow t \bar t) 
\\
&&  - {\pi \over 16 \hat s^2 } \biggl \lbrack 3 |C(x_s)|^2 
- {2\alpha_s }{{\rm Re}[C(x_s)] \over (M_t^2-\hat t)(M_t^2-\hat u)} 
\biggr\rbrack \nonumber \\
&& \times\biggl\lbrack 6 M_t^8-4 M_t^6(\hat t+\hat u) 
+4M_t^2 \hat t \hat u 
(\hat t+ \hat u) -\hat t\hat u(\hat t^2 + \hat u^2) + M_t^4(\hat t^2 -6 
\hat t\hat u +\hat u^2)\biggr\rbrack ~,
\nonumber 
\end{eqnarray}
where $x_s \equiv \frac{\sqrt{\hat s}}{M_S}$ in the ADD model, and $x_s
\equiv \frac{\sqrt{\hat s}}{m_0}$ in the RS model. The remaining
model-dependence is absorbed into the function $C(x)$, which is defined as
\begin{eqnarray}
{\rm ADD:}~~C_{ADD}(x) & = & \frac{16\pi}{M_S^4} \lambda_{ADD}(x) 
\nonumber \\
{\rm RS:}~~C_{RS}(x) & = & \frac{32 \pi c_0^2}{m_0^4} \lambda_{RS}(x)
\label{e1}
\end{eqnarray}
where
\begin{eqnarray}
\lambda_{ADD}(x_s) & = & 
\kappa^2 \sum_{\vec{n}} \frac{1}{\hat s - M_{\vec{n}}^2 + i\epsilon} 
\nonumber \\ 
\lambda_{RS}(x_s)  & = & 
m_0^2 \sum_n \frac{1}{\hat s - M_n^2 + iM_n \Gamma_n} \ .
\end{eqnarray}
where the $M_n$ are the masses of the individual resonances and the
$\Gamma_n$ are the corresponding widths. In the ADD model, the individual
graviton KK modes decay only through the feeble gravitational constant
$\kappa = \frac{2}{\mpl}$ and hence they are assumed to be stable, at
least for the purpose of collider studies. In the RS model, the graviton
widths are obtained by calculating their decays into final states
involving SM particles. This gives
\begin{equation} 
\Gamma_n = m_0 c_0^2 x_n^3 \Delta_n
\end{equation} 
where 
\begin{equation}
\Delta_n = \Delta_n^{\gamma \gamma} + \Delta_n^{gg} 
         + \Delta_n^{WW} + \Delta_n^{ZZ} 
         + \sum_\nu \Delta_n^{\nu\nu} + \sum_l \Delta_n^{ll} 
         + \sum_q \Delta_n^{qq} 
         + \Delta_n^{HH} 
\end{equation}
and each $\Delta_n^{aa}$ is a numerical coefficient arising in the decay
$h^n \to a \bar a$. For the partial width $\Delta_n^{HH}$, we have
fixed $M_H = 250$ GeV in our numerical studies. However, we have checked 
that variation of $M_H$ in the region allowed by fits to electroweak
precision data affects the results presented here only very weakly.  

In defining $C(x)$ and $\lambda(x)$ in the above fashion, we are motivated
by the connection with earlier studies of virtual effects of graviton
exchange in the ADD model. In these studies of the ADD model, $C(x) \sim
\frac{\lambda}{M_S^4}$ (where $M_S$ is the string-scale) and it has been
customary to assume $\lambda$ to be an energy-independent constant of
(${\cal O}(1)$). However, even in the ADD model, $\lambda(x)$ is constant
only in the limit $x \ll 1$ for $d > 2$ and is slowly varying for $d = 2$
in the same limit. For $\hat s$ comparable with $M_S \sim 1$ TeV --- which
occurs at the Tevatron for a non-negligible fraction of the events ---
$x_s \sim {\cal O}(1)$, and then $\lambda(x_s)$ shows appreciable
variation with $x_s$. For the RS model, where $m_0 \sim 100$ GeV is
possible, we are never in the $x_s \to 0$ regime for $t \bar t$ production
at the Tevatron and it is imperative to consider its variation. The
low-energy approximation becomes even worse at the LHC, unless we choose
the scales $M_S$ and $m_0$ to be very high \footnote{This would defeat the
spirit of theories with low-energy quantum gravity.}.

Calculation of $\lambda_{ADD}(x_s)$ is done by assuming the graviton KK
spectrum to form a quasi-continuum and then integrating over the states,
after having defined a suitable density-of-states function. This has been
done explicitly in Refs.~\cite{grw} and \cite{hlz}, and an elegant set of
formulae are presented in Appendix A of Ref.~\cite{hlz}.  Using their
results, one can account for the full-energy dependence of
$\lambda_{ADD}(x_s)$ and this turns out to be dependent on the number $d$
of extra dimensions as well as the energy.  It is, therefore, worthwhile to
include the full energy dependence of $\lambda_{ADD}(x_s)$ and redo the
analysis for the ADD case for both the Tevatron and the LHC.

For the RS model, given the masses and the widths of the individual
graviton resonances, we have to sum over all the resonances to get the
value of $\lambda_{RS}(x_s)$. We perform this sum numerically, using the
fact that the higher zeros of the Bessel function become evenly-spaced.
The summation is somewhat tricky because we have to sum over {\it several}
resonances. The heavier resonances are rather wide and tend to overlap. In
our numerical procedure, for a given value of $x_s = \frac{\sqrt{s}}{m_0}$, 
we retain all resonances which contribute with a
significance greater than one per mil, and treat the remaining KK modes as
virtual particles (in which case the sum can be done analytically).

We can now compute the integrated $ t \bar t$ production cross-section at
the Tevatron and the LHC, using the equation
\begin{equation}
\sigma (AB \rightarrow t \bar t)= \sum_{a,b} \int dx_1
dx_2 ~d\hat t ~\lbrack f_{a/A}(x_1)~ f_{b/B}(x_2) + x_1 \leftrightarrow x_2 
\rbrack ~{d\hat \sigma \over d\hat t}(ab \to t \bar t) ~,
\label{e5}
\end{equation}
where $A,\ B$ are the initial hadrons (either $p \bar p$ or $pp$), and
$f_{a(b)/A}$ denotes the probability of finding a parton $a$($b$ ) in the
hadron $A$ (similarly $f_{a(b)/B}$). The sum in Eq.~\ref{e5} runs over the
contributing sub-processes. 

The sub-process cross sections for the new physics that we have calculated
are at leading order. For the SM contribution to $t \bar t$ production in
hadronic collisions, significant progress has been made in computing
higher-order corrections. Not only have the next-to-leading order
corrections been calculated a long time ago\cite{nason}, but the
resummation of soft gluons and its effect on the total cross-section have
been computed\cite{catani}. In principle, a reliable estimate of the
cross-section for the case under consideration can also be made only when
we have (at least) the corrections to these processes at next-to-leading
order. For want of such a calculation, however, the best we can do is to
use the leading order QCD cross-section and the resummed QCD
cross-sections\cite{catani} to extract a `$K$-factor'. We work within the
approximation that the new physics will also be affected by QCD
corrections in a similar fashion so that we can fold in our cross-sections
for this case by the {\it same} $K$-factor. This is clearly an
approximation but it turns out to affect the constraints that we extract
from the data only marginally.

For our numerical studies, the above formulae have been incorporated into
a simple parton-level Monte Carlo event generator, where we have used the
CTEQ4M parametrisations\cite{cteq} for the parton distributions. Since it
is only the total $t \bar t$ cross-section that is being compared, we
expect the results from a parton-level study to be of reasonably high
quality. In order, now, to determine the constraints on the relevant
model, we simply compare the new physics contributions with the maximum
allowed deviation (at 95\% confidence level) from the SM cross-section
which is allowed by the experimental determination of the total $t \bar t$
cross-section in $p \bar p$ collisions at 1.8 TeV at the Tevatron. The
latter has been estimated to be $(6.5^{+1.7}_{-1.4})$ pb from 109 pb$^{-1}$ of
data collected by the CDF Collaboration \cite{cdf}, and to be $(5.9 \pm
1.7)$ pb from 125 pb$^{-1}$ of data collected by the D0
Collaboration \cite{d0}. These results are, of course, consistent at the
$1\sigma$ level, but the slight difference in errors estimated by the two
collaborations, in turn, slightly changes the bounds on the model
parameters derived in the two cases.

We first present our improved results on the string scale $M_S$ for the
case of the ADD model. In Table 1, we have listed, for different numbers
of large extra dimensions, the lower bounds on $M_S$ arising out of the
experimental numbers at Tun I of the Tevatron, as well as estimates for
the discovery limits at RUN II and at the LHC.

%---------------------------------------------------------------------------
\begin{center}
$$
\begin{array}{|c|cc|c|c|}
\hline
d & {\rm CDF~(GeV)} & {\rm D0~(GeV)} & {\rm Run~II~(GeV)} & {\rm LHC~(GeV)}\\
\hline\hline
2 &   1890~(1963) &   1984 (2088)  &  2678~(3228) &  6947~(13411) \\ 
3 &   1650~(1625) &   1722~(1699)  &  2272~(2252) &  6234~(6061) \\ 
4 &   1516~(1367) &   1575~(1428)  &  2049~(1888) &  6068~(4956) \\ 
5 &   1428~(1235) &   1478~(1291)  &  1904~(1705) &  5883~(4454) \\ 
6 &   1366~(1149) &   1411~(1202)  &  1800~(1586) &  5779~(4127) \\ 
\hline\hline
\end{array}
$$
\end{center}
\noindent {\bf Table 1}: {\footnotesize\sl 95\% C.L. lower bounds on the
string scale in the ADD model derived from the $t \bar t$ cross-section,
for different numbers $d$ of large extra dimensions. The first two columns
show actual bounds derivable from existing data from Run-I of the
Tevatron, while the last two columns show estimates of the discovery
limits from Run-II of the Tevatron and from the LHC. Numbers in
parentheses show those that would have been obtained by ignoring the
energy dependence of $\lambda_{ADD}(x_s)$, {\it i.e.} by setting 
$\lambda(x_s)$ to its $x_s \ll 1$ limit irrespective of $\sqrt{\hat s}$.}
%---------------------------------------------------------------------------
\vskip 5pt

In order to make an estimate of the discovery limits on $M_S$ obtainable
from Run-II of the Tevatron, we assume that the $p \bar p$ collisions take
place at a centre-of-mass energy of 2 TeV, and (conservatively) 
that an integrated
luminosity of 2 fb$^{-1}$ will be collected. Taking improvements in 
the $b$-tagging efficiency into account, this has been
estimated \cite{Run-II} to lead to an uncertainty in the cross-section 
of 8-10\%,
which can be translated, using the SM cross-section, into an error bar of
about 0.7 pb.
Since this is significantly smaller than the errors from Run-I, we 
predict significantly better results from Run-II of the Tevatron, assuming
that the experimental central value will be very close to the SM prediction.

It may appear that the bounds shown in Table~1 are considerably better
than the bounds on the effective string scale $M_S/\lambda^{1/4}$ earlier
derived \cite{ours1}, which were in the range 600--700 GeV. There is no
paradox, involved, however, since the value of $\lambda$ is not actually
${\cal O}$(1), but is more like $16\pi$. Accordingly, the fourth root
provides a factor somewhat larger than 2, which accounts for the numbers
being in the present ballpark. It is also obvious that the low-energy
approximation for $C(x_s)$ is not a very good one, except perhaps, for
$d = 3$. 

Going from the Tevatron to the LHC ($pp$ collisions at $\sqrt{s} =
14$~TeV) affects the results quite significantly. This is because of the
higher energy, and the fact that the $gg$ channel is dominant at the LHC
energy. We recall that the interference between the SM and the ADD model
is present only in the $gg$ channel. In performing the ADD analysis for
the LHC, we encounter the technical problem that there is a significant
part of the phase-space for which $x_s > 1$, which means that the
low-energy effective theory approach becomes meaningless. In order to
handle this problem we have used kinematic cuts on $\sqrt{\hat s}$ to make
sure that $x_s$ remains in the acceptable range. To derive bounds, we
assume that no new physics effect will be seen, in which case it is likely
that the experimental central value of the cross-section will coincide with
the SM prediction, about 868 pb --- which means that we can expect more
than eight-and-a-half million $t \bar t$ events at the LHC, for a
projected (integrated) luminosity\footnote{This is actually a conservative
estimate: estimates of around 100~fb$^{-1}$ have also been made.} of
10~fb$^{-1}$. The statistical error is, therefore, negligibly small and
the experimental error is expected to be dominated by the systematics. We
assume a systematic error of 5 pb in the cross-section for the results
shown in Table 1, which may be regarded as an educated guess. We have
verified that reasonable changes in this number do not affect the
discovery limits very significantly, since the cross-section has a generic
$M_S^{-4}$ dependence.

It is interesting to note that the low-energy approximation for $M_S$
becomes wildly inaccurate for the case of LHC for the case $d = 2$. 
The actual lower bound
on $M_S$, when the energy dependence is taken into account, is much more
modest, and shows smaller variation with the number of extra dimensions,
(about 20\%, as we go from $d = 2$ to $d = 6$) than do the Tevatron bounds
(about 35\%, as we go from $d = 2$ to $d = 6$). 

% ------------------------------------------------------------------
\begin{figure}[htb]
\begin{center}
\vspace*{3.6in}
      \relax\noindent\hskip -4.2in\relax{\includegraphics{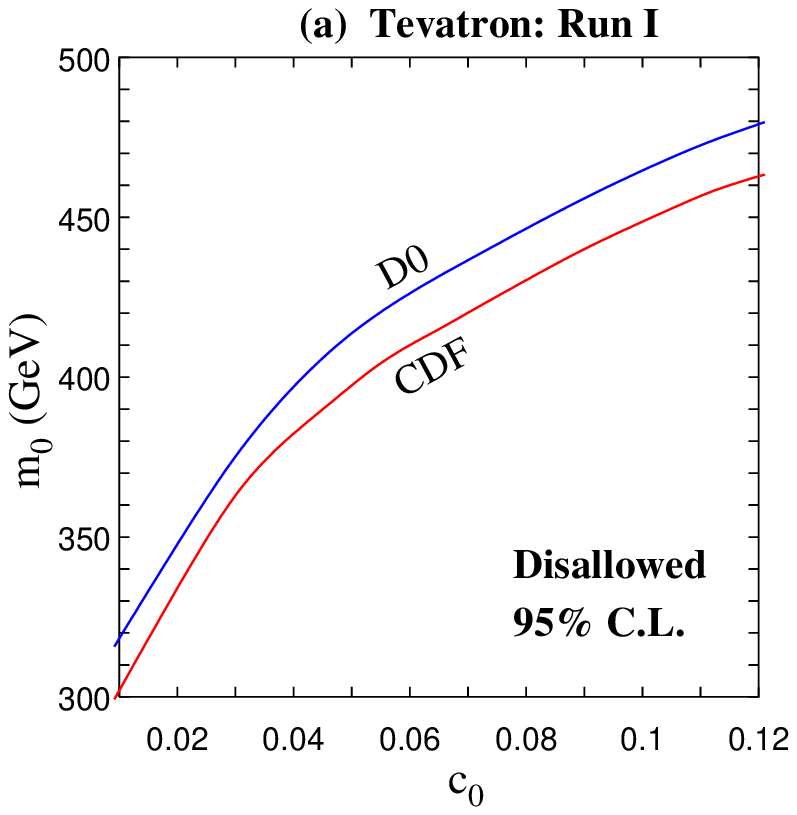}}
                     \hskip  3.2in\relax{\includegraphics{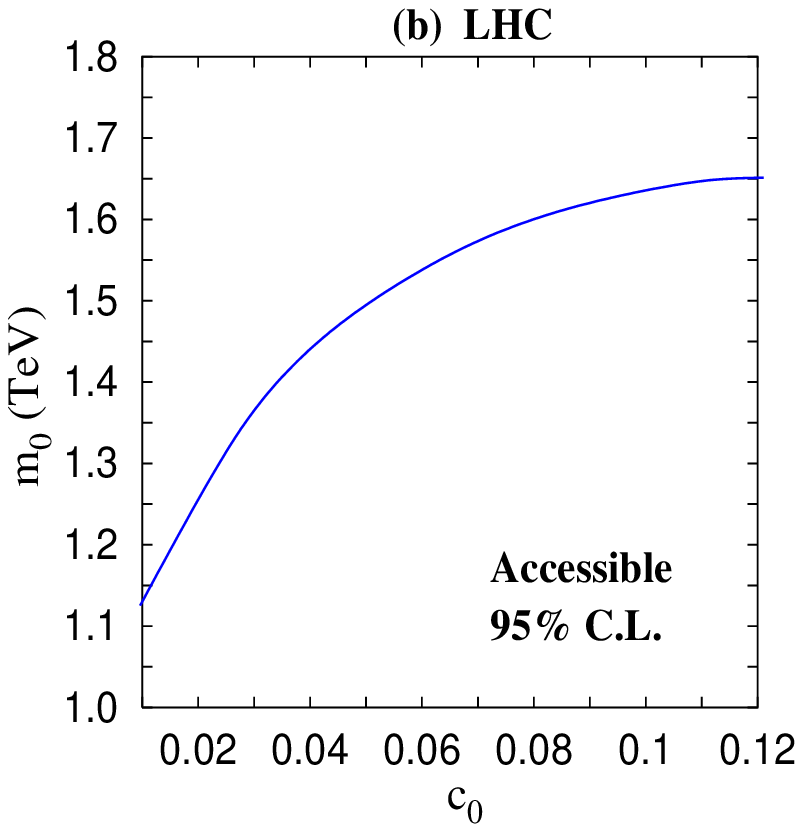}}
\end{center}
\end{figure}
\vspace*{-1.0in}
\noindent {\bf Figure 1}.
{\footnotesize\it Illustrating ($a$)
95\% C.L. constraints on the $m_0-c_0$ plane of the Randall-Sundrum model,
using $t \bar t$ production data from Run-I of the Tevatron, and ($b$)
discovery limits at 95\% C.L. on the $m_0-c_0$ plane at the LHC.  }
% ------------------------------------------------------------------
\vskip 5pt

We now turn to the more interesting case of the RS model, where the
contribution from the $t \bar t$ cross-section have not been studied
before. In this case, as we have seen, the parameter space is
two-dimensional, with both $m_0$ and $c_0$ being free parameters within
certain theoretical constraints.  We present the results of our
computations in Fig.~1. In Fig~1($a$), we show the excluded region at 95\%
C.L. in the $(m_0, c_0)$ plane obtained by comparing the excess predicted
by the RS model with the experimental errors from the CDF and D0
experiments, respectively. It may be seen that the present data already
constrain a significant region of the parameter space and for the values
of $c_0$ between 0.01 and 0.1, suggested by naturalness considerations
discussed earlier, $m_0$ values smaller than about 300~GeV (rising to
about 500~GeV for larger values of $c_0$) are definitely excluded. This
means that the first KK graviton resonance must lie above 1 TeV at the
least. As in the case of the ADD model, the constraints are expected to
get much better at Tevatron Run II, unless, indeed, a discovery is made.
To underline the importance of understanding the error estimates at the 
Tevatron better, we have not presented discovery limits for the Tevatron,
since the RS model case is more sensitive to small changes in the error 
estimate than the ADD model. 

At the LHC, our analysis of the RS model lies on a firmer theoretical
foundation, since the low-energy theory is valid for the entire energy
range accessible at this machine. However, our numerical analysis must
take care of more graviton resonances, which can become accessible because
of the larger energy. Once this is done, we assume the same errors of 5 pb
as before, and obtain discovery limits in the ($m_0$,$c_0$)-plane as in
the case for the Tevatron. These are exhibited in Fig.~1($b$). It is
obvious that the accessible values of $m_0$ for $0.01<c_0<0.1$ lie in the
range between 1.1~TeV and 1.7~TeV, which means that the first graviton
resonance can be probed up to around 4~TeV. These numbers change somewhat
as the systematic error is varied from 5 pb, but so long as the latter is
not changed drastically, the discovery limits remain in the range
1--2~TeV. Improving the systematic error to 1 pb or less, would, however,
drive the discovery limits well above 2~TeV, and this may be a goal to set
for experimental analyses, when the time is ripe to make serious
estimates.

The fact that the LHC can probe KK graviton resonances of the RS model in
the range of a few TeV is a rather exciting prospect. This is because the
graviton resonances cannot be made indefinitely heavy in this model.  For,
if $m_0$ becomes very large, it would require either ${\cal K}$ to be
large, or ${\cal K}R_c$ to be small (see Eq.~\ref{m0c0}). In either case,
the curvature of the fifth dimension becomes very large, and this leads to
problems in fine-tuning the cosmological constants to get a flat metric on
the 3-branes. Moreover, a small warp factor $e^{-\pi{\cal K}R_c}$ would
tend to increase the electroweak scale, leading to unitarity problems in
the Higgs sector; this can be ameliorated by reducing the scale on the
Planck brane which gives rise to the electroweak symmetry-breaking, but
only at the cost of introducing a small hierarchy. In the simplest version
of the RS model, therefore, we expect the graviton spectrum to start at a
few TeV, and this, as we have shown, is precisely the range which can be
probed at the LHC in $t \bar t$ collisions. The situation is different in
the ADD model, since the string scale $M_S$ can easily be pushed above the
reach of the LHC without encountering any theoretical difficulties.

To summarize, we have analysed the effects of the interactions of the
spin-2 Kaluza-Klein modes with SM fields in $t \bar t$ production at the
Tevatron and LHC, both in the case of large extra dimensions, as suggested
by ADD, as well as for a small warped extra dimension, as suggested by RS.
Using the full dependence of the effective coupling of gravitons to SM
particles on the parton-level centre-of-mass energy, we have re-derived
95\% C.L. bounds from Run I data at the Tevatron on the string scale $M_S$
in the ADD model. We find that these vary from about 1.3 to 2.0 TeV,
depending on the number of large extra dimensions.  For the RS model, the
respective bound on the mass scale $m_0$ of graviton resonances varies
from around 300--500~GeV, depending on the coupling parameter $c_0$. These
numbers can improve at Run-II of the Tevatron, but it is only at the LHC
where we can expect the discovery limits to go up by about three times the
present lower bounds. In particular, we expect that $M_S$ values in the
range 5.7~TeV to 6.9~TeV for the ADD model and $m_0$ in the range 1.1~TeV
to 1.7~TeV for the RS model can be probed in $t \bar{t}$ production at the
LHC. This shows that $t \bar{t}$ hadroproduction is a powerful tool to
probe extra dimensions, and we may expect very interesting results to come
out of such measurements in the near future.

\vskip20pt
{\sl Acknowledgements:} 
PM was supported by FAPESP (Processo:99/05310-9).  He would also like to
thank the High Energy Section of the Abdus Salam ICTP for its hospitality
during the early stages of this work.  SR and KS would like to thank the
Theory Division of CERN for hospitality.

%-------------------------------------------------------------------------
 
%-------------------------------------------------------------------------

\begin{thebibliography}{999} 
 
\bibitem{string}
P.~Horava and E.~Witten, {\it Nucl. Phys.} {\bf B460} (1996) 506;
J. Polchinski, {\it Lectures given at the 
SLAC Summer Institute: Gravity -- From the Hubble Length to the 
Planck Length (SSI 98), Stanford, 3-14 Aug 1998}, hep-th/9812104;
J.D.~Lykken {\it Phys. Rev.} {\bf D54} (1996) 3697;
E.~Witten, {\it Nucl. Phys.} {\bf B471} (1996) 135.

\bibitem{dimo} 
N.~Arkani-Hamed, S.~Dimopoulos and G.~Dvali, 
{\it Phys. Lett.} {\bf B249} (1998) 263;
I.~Antoniadis, N.~Arkani-Hamed, S.~Dimopoulos and G.~Dvali, 
{\it Phys. Lett.}  {\bf B436} (1998) 257. 

\bibitem{gravexp} J.~C.~Long, H.~W.~Chan and J.~C.~Price, 
{\it Nucl. Phys.}  {\bf B539} (1999) 23. 

\bibitem{dimo4} N. Arkani-Hamed, S. Dimopoulos and G. Dvali, 
{\it Phys. Rev.}  {\bf D59} (1999) 086004. 

\bibitem{phenoadd} For a review of the phenomenology of the ADD
model, see K. Sridhar, {\it Int. J. Mod. Phys} {\bf A15}, (2000) 2397 
(hep-ph/0004053); S. Raychaudhuri, Talk given at the Sixth Workshop
on High-Energy Phenomenology, Chennai (India), January 2000. For
details of the phenomenology see Ref.~\cite{ours}.

\bibitem{cullen} S. Cullen and M.~Perelstein, 
{\it Phys. Rev. Lett.} {\bf 83} (1999) 268.
 
\bibitem{rs} L. Randall and R. Sundrum, {\it Phys. Rev. Lett.} {\bf 83}
(1999) 3370.

\bibitem{csaki} C. Csaki, M. Graesser, L. Randall and J. Terning,
{\it Phys. Rev.} {\bf D 62} (1999) 045015, C. Csaki, M. Graesser and
G.D. Kribbs, Santa Cruz Preprint No. SCIPP-00-27, 
hep-th/0008151 (2000).

\bibitem{gold} W.D. Goldberger and M.B. Wise, Phys. Rev.  Lett. 
83 (1999) 4922; {\it Phys.Lett.} {\bf B475}(2000) 275.

\bibitem{bagger} R. Altendorfer, J. Bagger and D. Nemeschansky,
CITUSC-00-015, hep-th/0003117.

\bibitem{radion} U. Mahanta and A. Datta, {\it Phys.Lett.} {\bf B483}
(2000) 196; U. Mahanta, Mehta Research Institute Preprint (2000) 
hep-ph/0008042; K. Cheung (2000) hep-ph/0009232.  

\bibitem{verlinde} Chang S. Chan, Percy L. Paul and H. Verlinde,
{\it Nucl. Phys.} {\bf B581} (2000) 156.

\bibitem{grw} 
G.~F.~ Giudice, R.~Rattazzi and J.~D.~Wells, 
{\it Nucl. Phys.}  {\bf B544} (1999) 3; 

\bibitem{hlz} 
T.~Han, J.~D.~Lykken and R-J.~Zhang, {\it Phys. Rev.}  {\bf D59} (1999) 105006.
(For the corrected formulae refer to hep-ph/9811350 v.4). 

\bibitem{dhr} H. Davoudiasl, J.L. Hewett and T.G. Rizzo, 
{\it Phys. Rev. Lett} {\bf 84} (2000) 2080; SLAC Preprint SLAC-PUB-8436 (2000) 
hep-ph/0006041.

\bibitem{drs} P. Das, S. Raychaudhuri and S. Sarkar, {\it JHEP} {\bf 0007:050}
(2000). 

\bibitem{gr} D. Ghosh and S. Raychaudhuri, TIFR Preprint TIFR-TH-00-38,
hep-ph/0007534.

\bibitem{giddings} S.B. Giddings and E. Katz, MIT Preprint No. MIT-CTP-3024, 
hep-th/0009176 (2000).

\bibitem{hewett} 
J.~L.~Hewett, SLAC Preprint SLAC-PUB-8001, hep-ph/9811356 (1998).

\bibitem{ours1}
Prakash Mathews, Sreerup Raychaudhuri and K.~Sridhar, {\it Phys. Lett.} 
{\bf B 450} (1999) 343.

\bibitem{nason} P.~Nason, S.~Dawson and R.K.~Ellis,
{\it Nucl. Phys.} {\bf B303}, 607 (1988);
{\it Nucl. Phys.} {\bf B327}, 49 (1989).

\bibitem{catani} S.~Catani et al., {\it Phys. Lett.} {B 378} (1996) 329;
{\it Nucl. Phys.} {\bf B 478}, (1996) 273.
 
\bibitem{cteq} 
H.L.~Lai et al., {\it Phys. Rev.} {\bf D51}, 4763 (1995).

\bibitem{cdf} S. Leone, in the Proceedings XXXVth Rencontres de Moriond: 
Electroweak and Unified Theories, Les Arcs, France, (2000)
FERMILAB-CONF-00/115-E. 

\bibitem{d0} B.~Abbott et al.,
{\it Phys. Rev. Lett.} {\bf 80}, 666 (1998). For the latest cross-scetion
values refer to the talk of S.Leone quoted above.

\bibitem{Run-II} 
P.C. Bhat , H. Prosper  and Scott S. Snyder, 
{\it Int.J.Mod.Phys.} {\bf A13} (1998) 5113.

\bibitem{ours}
Prakash Mathews, Sreerup Raychaudhuri and K.~Sridhar, {\it Phys. Lett.} 
{\bf B 450} (1999) 343; {\it Phys. Lett.} {\bf B 455} (1999) 115;
{\it JHEP} {\bf 0007:008} (2000). 

\end{thebibliography}
\end{document}